\documentclass[journal=jacsat,manuscript=article]{achemso}
\SectionNumbersOn  

\makeatletter
\renewcommand*{\acs@author@fnsymbol}[1]{\textsuperscript{\number#1}}
\makeatother
\usepackage{etoolbox}
\makeatletter
\patchcmd{\acs@contact@details}{E}{*\,E}{}{}
\makeatother
\makeatletter
\renewcommand*\acs@author@fnsymbol[1]{%
  \textsuperscript{%
    \ifnum#1=0
      *
    \else
      \number#1
    \fi
  }%
}
\makeatother

\usepackage{etoolbox}
\makeatletter
\patchcmd{\acs@contact@setup}
  {\footnotemark[0]}
  {}{}{}

\patchcmd{\acs@author@format}
  {\textsuperscript{\acs@author@fnsymbol{1}}}
  {\textsuperscript{\acs@author@fnsymbol{1}}\textsuperscript{*}}
  {}{}
\makeatother

\usepackage[version=3]{mhchem} 
\usepackage{float}
\usepackage{amssymb}

\author{Rébecca Loubet}
\affiliation[Laboratory of Engineering Thermodynamics (LTD)]
{Laboratory of Engineering Thermodynamics (LTD), RPTU Kaiserslautern, Germany}
\author{Pascal Zittlau}
\affiliation[Laboratory of Engineering Thermodynamics (LTD)]
{Laboratory of Engineering Thermodynamics (LTD), RPTU Kaiserslautern, Germany}
\author{Marco Hoffmann}
\affiliation[Laboratory of Engineering Thermodynamics (LTD)]
{Laboratory of Engineering Thermodynamics (LTD), RPTU Kaiserslautern, Germany}
\author{Luisa Vollmer}
\affiliation[Visual Information Analysis Research Group (VIA)]
{Visual Information Analysis Research Group (VIA), RPTU Kaiserslautern, Germany}
\author{Sophie Fellenz}
\affiliation[Machine Learning Research Group (ML)]
{Machine Learning Research Group (ML), RPTU Kaiserslautern, Germany}
\author{Heike Leitte}
\affiliation[Visual Information Analysis Research Group (VIA)]
{Visual Information Analysis Research Group (VIA), RPTU Kaiserslautern, Germany}
\author{Fabian Jirasek}
\affiliation[Laboratory of Engineering Thermodynamics (LTD)]
{Laboratory of Engineering Thermodynamics (LTD), RPTU Kaiserslautern, Germany}
\author{Johannes Lenhard}
\affiliation[Laboratory of Engineering Thermodynamics (LTD)]
{Laboratory of Engineering Thermodynamics (LTD), RPTU Kaiserslautern, Germany}
\alsoaffiliation[Philosophy in Science in Engineering]
{Philosophy in Science and Engineering, RPTU Kaierslautern, Germany}
\author{Hans Hasse}
\affiliation[Laboratory of Engineering Thermodynamics (LTD)]
{Laboratory of Engineering Thermodynamics (LTD), RPTU Kaiserslautern, Germany}
\email{hans.hasse@rptu.de}
\phone{+49(0)631 205-3497}
\fax{+49(0)631 205-3835}

\title[An \textsf{achemso} demo]
  {Superstudent intelligence in thermodynamics}

\abbreviations{LLM}

\begin{document}

\begin{tocentry}




  \centering
  \includegraphics[width=9.5cm,height=4.12cm,keepaspectratio]{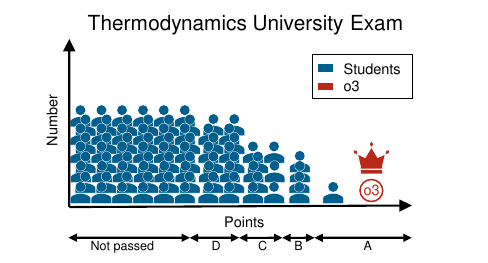}


\end{tocentry}

\begin{abstract}
In this short note, we report and analyze a striking event: OpenAI's large language model o3 has outwitted all students in a university exam on thermodynamics. The thermodynamics exam is a difficult hurdle for most students, where they must show that they have mastered the fundamentals of this important topic. Consequently, the failure rates are very high, A-grades are rare - and they are considered proof of the students' exceptional intellectual abilities. This is because pattern learning does not help in the exam. The problems can only be solved by knowledgeably and creatively combining principles of thermodynamics. We have given our latest thermodynamics exam not only to the students but also to OpenAI’s most powerful reasoning model, o3, and have assessed the answers of o3 exactly the same way as those of the students. In zero-shot mode, the model o3 solved all problems correctly, better than all students who took the exam; its overall score was in the range of the best scores we have seen in more than 10,000 similar exams since 1985.  This is a turning point: machines now excel in complex tasks, usually taken as proof of human intellectual capabilities. We discuss the consequences this has for the work of engineers and the education of future engineers.
\end{abstract}


\section{Background: LLMs solve thermodynamic problems}
\label{section:background_llms}

Large language models (LLMs) are developing at a breathtaking pace. We started monitoring their performance in solving thermodynamic problems about a year ago. The results were published in \citeauthor{Loubet2025} (\citeyear{Loubet2025}) \cite{Loubet2025} and are therefore only briefly summarized here. In all studies, including the one we report here, zero-shot prompting was used, i.e., the LLMs were taken off the shelf and used without any special training. Two sets of textbook-style problems were used for the tests, a simple and an advanced one. The ones from the former set were far simpler than those we give to our students in the thermodynamics exam (see Section~\ref{section:exam_and_results}). The ones from the latter set were closer to exam problems. However, all problems were tuned for LLMs, e.g., by removing graphical input and not asking questions that require graphical output, such as diagrams. The following LLMs were used: GPT-3.5, GPT-4, and GPT-4o of OpenAI, Llama 3.1 70B of Meta, and le Chat (Mistral Large 2) of MistralAI. Table~\ref{tbl:scores of llms} gives an overview of their scores on the two sets. We will not repeat the discussion from \citeauthor{Loubet2025} (\citeyear{Loubet2025}) \cite{Loubet2025} here and only point out that there was a significant improvement going from GPT-3.5 to GPT-4 and GPT-4o but that in late 2024, when we carried out the test, none of the models was able to satisfactorily solve the advanced problems. 
\begin{table}
  \caption{Scores of different LLMs in solving thermodynamic problems from different test sets. Results from \citeauthor{Loubet2025} (\citeyear{Loubet2025}) \cite{Loubet2025}. The number reported for each model is the average percentage of the maximal achievable points obtained in the test. }
  \label{tbl:scores of llms}
  \begin{tabular}{|r|c|c|c|c|c|}
    \hline
    & \multicolumn{5}{c|}{LLM}  \\ [1ex]
    \hline
    Test set   & GPT-3.5 & GPT-4 & GTP-4o & Llama 3.1 70B & Le Chat \\ [1ex]
    \hline
    Simple & 45.8 \% & 88.6 \% & 87.1 \% & 75.9 \% & 72.8 \% \\
    \hline
    Advanced  & n.a. & 47.5 \% & 55.2 \% & 40.7 \% & 51.9 \%  \\
    \hline
  \end{tabular}
\end{table}

A few months later, in April 2025, we extended the study of the performance of LLMs in thermodynamics to the model o3 from OpenAI \cite{o3_OpenAI}. However, the test differed from the previous ones: we did not use the problems specifically tailored for LLMs. We simply took the exam that we had given to our students a few days earlier and gave it to o3, which had been released just a few days earlier. The exam also included graphical input and questions, for which graphical output was required. The outcome is breathtaking, as shown in the following section.

\section{The exam and its results}
\label{section:exam_and_results}
The exam we refer to here belongs to our first course in engineering thermodynamics. 90 students took the exam this year. As we gave the same problems to the o3 model and our students and evaluated the results in the same way, we can directly compare the performance of the o3 model with that of our students. The exam and a solution are provided in the Supplementary Information. The test was carried out in German, but we know from our previous studies that the language has hardly any influence on the performance of LLMs in solving thermodynamic problems - at least as far as English and German are concerned. No special prompt engineering was used; the prompt was chosen based on some preliminary tests. In an English translation, the zero-shot prompt was: 
\begin{quote}
\textit{“You are an expert in thermodynamics. Solve the exam given below. There are three problems 1-3. In each of them, start by answering the first question a). If you are asked for plots, draw them.}
\end{quote}

After this, the text of Problem 1 was prompted. After receiving the answer, the text for Problem 2 was prompted. Finally, the same was true for Problem 3. 
Three independent runs were carried out with o3. In contrast to our previous studies\cite{Loubet2025}, the results hardly differed, but we still report the three scores individually.  

\begin{figure}[!htbp]
\centering
\includegraphics[width=0.84\textwidth]{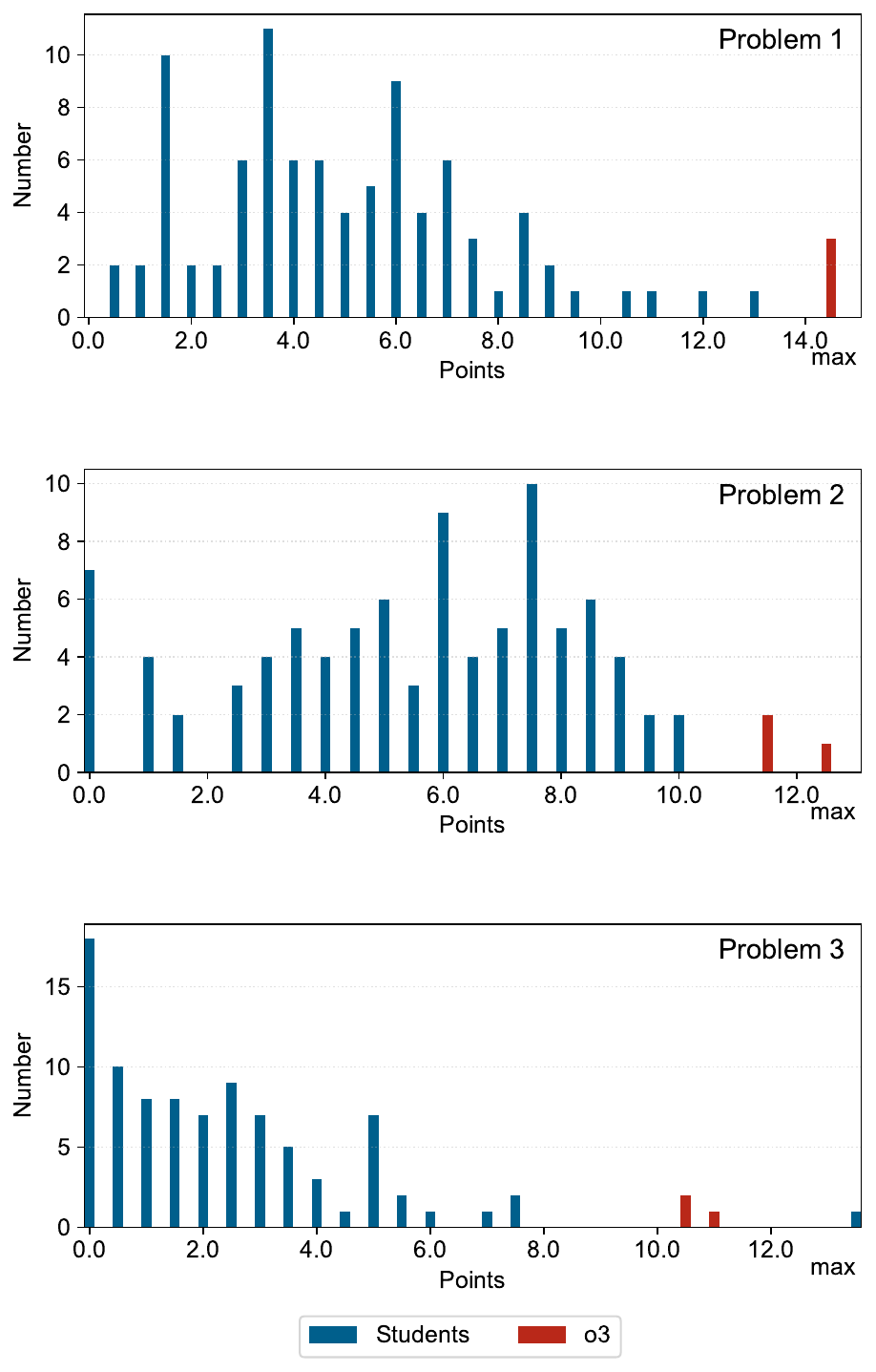}
\caption{Distribution of points earned by o3 (three runs) and by 90 students in an exam in engineering thermodynamics in spring 2025. The three panels show the scores obtained in the three problems of the exam.}
\label{fig.results problems}
\end{figure}

\begin{figure}[!htbp]
\centering
\includegraphics[width=0.84\textwidth]{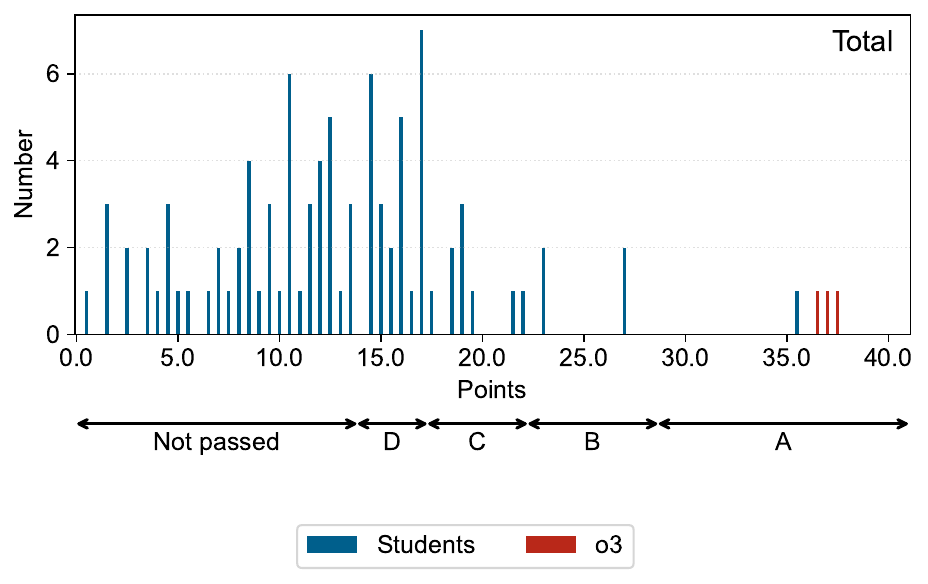}
\caption{Distribution of points earned by o3 (three runs) and by 90 students in an exam in engineering thermodynamics in spring 2025. The panel shows the overall score with resulting grades.}
\label{fig.results total}
\end{figure}

The results from o3 and those of the students are compared in Figures~\ref{fig.results problems} and~\ref{fig.results total} - the performance of o3 is impressive. Model o3 basically solved the first problem without errors. The same holds for the second problem, where points were only lost because of minor issues of o3 with graphical representations. The only major mistake o3 made was in Problem 3, which relates to a fairly special topic: the constant-density fluid's caloric properties.

\section{Discussion}
\label{section Discussion}

The exam discussed here is that of a course in which the fundamental concepts of thermodynamics are introduced, including the first and second law, as well as fundamentals of describing thermodynamic properties of pure components. The corresponding author of this report has taught this course for several decades to more than 10,000 students at different universities. For most students, thermodynamics is a difficult hurdle; the failure rates in the exam are very high, and grades A are extremely rare. A grade A in thermodynamics is often considered indisputable proof of the exceptional intelligence of the student. This is because the problems in the thermodynamics exam are always new and carefully crafted, so they cannot be solved by relying on pattern learning but only by knowledgeably and creatively applying the principles of thermodynamics. The purpose of the course is to teach this to students, and the exam is designed to determine whether this was successful. In evaluating the results, points are earned for analyzing the problem, formulating the applicable equations, and combining them. The correct numerical solution gives points, but only comparatively few, so that grade A can be earned, in principle, without doing any numerical calculations. We have always maintained these standards in our thermodynamics courses over the many years, and the students’ results in the exam shown in Figure~\ref{fig.results total} are typical. The high failure rate is regrettable but a consequence of maintaining high standards and being accepted in the German academic system. 

The results described above demonstrate that LLMs are developing at an amazing pace, and so is their ability to solve complex tasks. In our tests carried out about six months ago, in autumn 2024, LLMs could score above 90 \% only in simple thermodynamic problems, whereas they often scored below 50 \% in advanced problems, see Table~\ref{tbl:scores of llms}. In spring 2025, a reasoning LLM solves thermodynamic problems that are more demanding than the advanced problems from our recent study with scores above 90 \%, see Figures~\ref{fig.results problems} and~\ref{fig.results total}. 

In two recent studies from the literature, several LLMs were tested on problems from the Physics Olympiad\cite{ tschisgale2025evaluatinggptreasoningbasedlarge} and from
undergraduate physics\cite{xu2025ugphysicscomprehensivebenchmarkundergraduate}. In both cases OpenAI's reasoning models (o1-preview in Ref.\cite{ tschisgale2025evaluatinggptreasoningbasedlarge}
and o1-mini in Ref.\cite{xu2025ugphysicscomprehensivebenchmarkundergraduate})
performed best - however, not as good as o3 performed in our study. In Ref.\cite{ tschisgale2025evaluatinggptreasoningbasedlarge}, where the performance of the LLMs was also compared to those of human candidates, o1-preview outperformed average candidates but scored much worse than the best candidate. We consider this a turning point in academic education, at least in thermodynamics. Some consequences are discussed below. The substantial improvement of the results in our test compared to those in Refs. \cite{tschisgale2025evaluatinggptreasoningbasedlarge, xu2025ugphysicscomprehensivebenchmarkundergraduate} probably results from the improvements of the reasoning going from o1 to o3 rather than from the different nature of the studied problems. 

The findings from the present study also contribute to the ongoing discussion on the relation between knowledge-based models and data-driven models \cite{Jirasek_2023,Specht_2024,Meng_2025,Wu_2024,Todorovski_2006,Dubois_2000}. We have recently described a knowledge-based system for solving thermodynamic problems (\citeauthor{Vollmer_2024} (\citeyear{Vollmer_2024}) \cite{Vollmer_2024}), which has the advantage that the correctness of the solution can be guaranteed, which is not the case for LLMs. However, in its current form, this model has only a limited scope (similar to that of the simple problems from Table~\ref{tbl:scores of llms}), and setting up this model, even for this limited scope, was very tedious and so would be future extensions. The question of how to combine knowledge-based approaches with LLMs, e.g., for assessing their results, remains on the table, especially in science and engineering and in all fields where correctness guarantees for solutions are required. 

\section{A turning point and its consequences}
\label{section a turning point}

The results presented here are more than a substantial quantitative improvement; they mark a fundamental turning point. They show that machines now excel in complex tasks, which are usually taken as proof of human intellectual capabilities. In the following, we will therefore call such machines “intelligent”. Alan Turing made an analogous move in his famous paper\cite{TURING_1950} on the imitation game (now called the Turing test). There, Turing argues that machines that can chat with a human in such a way that the human cannot determine whether he is interacting with a machine or another human pass the test and should be considered intelligent. The thermodynamics exam presents a shortcut to the test: if something can excel in the exam, i.e., successfully imitate a very bright student, this is sufficient to count as being intelligent. For an extended discussion on the intelligence of machines, we refer the reader to Refs.\cite{Legg_2007,Chollet_05.11.2019,French_2000}. Bubeck et al. \cite{Bubeck_2023} have recently attested \textit{'sparks of artifical general intelligence'} to GPT-4 — we have found not only sparks but rather a blaze in our study of the distinctly more powerful OpenAI model o3.

The advent of intelligent machines has far-reaching consequences \cite{Eloundou_2023,Kasneci_2023,Messeri_2024,Binz_2025}. In this short paper, we focus on the consequences for engineering. Our ambition here is to contribute to the ongoing — and still early — discussion rather than to provide a comprehensive overview.

The way engineers work will change. We will see human engineers and intelligent machines (you may call them “AI colleagues”) working closely together in mixed teams. Our results suggest that this could happen on a large scale very soon. In such mixed teams, as in purely human teams, the big question is how to best assign the different tasks. We will have to learn which tasks to assign to intelligent machines and which to humans. Our results demonstrate that one of the problems will be the rapid development of the capabilities of intelligent machines. We may not have enough time to fully explore the pros and cons of specific ways of collaboration in mixed teams before new, more capable AI colleagues replace the current ones. The high fluctuation among AI colleagues could pose an extraordinary challenge for such mixed teams.  

Another critical question is which results from which team members can be trusted and which should be double-checked. This question is also not new, but it may become more difficult to answer as the capabilities of AI colleagues change continuously and rapidly. We only note that for the tasks we have studied here, the assessment of the LLMs' responses (though tedious because it was made by hand) was fairly straightforward and followed established routines, ensuring high confidence in the evaluation. This may be much more complicated in other situations.  

Decision-making and taking on responsibility are crucial in any team leadership. In mixed teams of humans and intelligent machines, challenging questions will arise here, such as: How should the results of intelligent machines be incorporated into team discussions? Which legal consequences has the incorporation of intelligent machines in workflows? Who will take the responsibility for decisions made based on their results? Do we always need to have humans in the decision process? 

Overall, the findings from our study indicate that intelligent machines will become important very rapidly and that it is high time to develop strategies for cooperating with them in the future — in engineering and beyond \cite{Smuha_2025}. 

All this also has significant consequences for the education of the next generation of engineers. We must decide which skills are still needed and which tasks can be routinely taken over by intelligent machines. The answer will be highly dynamic, and it will be interesting to see if the study and examination regulations of universities can keep pace with the current changes. The future engineers will have to be trained to cooperate with intelligent machines. This requires special skills related to the input and output of the intelligent machine. On the input side, prompting, learning to ask the right questions, and learning to ask them in the right way is essential. On the output side, we face one of the most challenging tasks of future engineers: they will have to decide whether or not to trust the results provided by the intelligent machine and, even more importantly, whether or not to apply them in a specific situation in practice. Making these decisions requires that engineers continue to master their discipline. However, the profile will shift: engineers will no longer need to excel in operational tasks — fundamental understanding will become more important than ever.   

Intelligent machines can also support education \cite{Chu_2025,Shojaei_2025,Supan_23.06.202412.07.2024,Fuligni_02.02.2025,Levonian_2025,Caines_2023,Kasneci_2023,Henkel_15.02.2024,Shen_02.06.2021,Lieb_2024,Vanzo_2024}. They could help students as a kind of private teacher – if (and only if) they are trustworthy. Checking this is not as new a question as it may seem. In principle, the same question is always on the table when selecting student teaching assistants — and, even more importantly, when hiring professors.
These current changes will also have to be taken into account for the academic exams. It is hard to motivate students to thoroughly prepare for an exam in which they know that an intelligent machine can easily outwit them. Rather than asking whether a student can perform certain tasks independently, the question may be whether the student has the fundamental understanding needed to produce reliable results in cooperation with intelligent machines. There is also a lively debate on using LLMs for grading exams \cite{Kortemeyer_2023,Kortemeyer_2024,Floden_2025}.

The developments we have described here may amplify inequalities in the engineering community: the performance of the best engineers will be boosted by cooperation with intelligent machines, and there will be fierce competition between companies to hire such engineers. Things may look less rosy at the other end of the distribution, which is more populated, see Figure 1. Intelligent machines can take over many routine engineering tasks, or these tasks can be organized so that much less human input is needed than currently. We acknowledge that an extrapolation from our study on academic exams to real engineering tasks is bold. LLMs are certainly better suited for solving the former than the latter. However, even only a year ago, none of us would have believed that an LLM would almost perfectly solve all the problems of our cherished thermodynamics exam. We should prepare to cooperate with highly skilled intelligent machines. 

Last but not least, the findings we have discussed here are also relevant to the epistemology of engineering. We mention only two aspects here. Firstly, we have always claimed that, with our exam, we test whether the student has understood thermodynamics. The exams are designed in such a way that we can find this out. Following this argument, we must now admit that the machine understands thermodynamics. This leads directly to the problem of what understanding means. The second aspect is related: The current prevailing opinion is that learning the principles of a discipline and some examples showing how to utilize them provide the means to tackle all other situations. That is the point of general principles. The question is whether the intelligent machine has learned the principles or profits mainly from seeing so many applications. Coming to grips with these questions is likely to become a major issue of the future epistemology of engineering.

\begin{acknowledgement}

The authors gratefully acknowledge financial support by \textsf{Deutsche Forschungsgemeinschaft (DFG)} in the frame of its \textsf{Priority Program 2331 Machine Learning in Chemical Engineering}.

\end{acknowledgement}

\bibliography{achemso-demo}

\end{document}


\section{Highlights}

\begin{itemize}
  \item OpenAI's new reasoning large language model o3 has outperformed 90 students in a university exam on thermodynamics in spring 2025.
  \item The thermodynamics exam is a high hurdle for most students, the failure rate was 58~\%. One student achieved a grade A — o3 performed better. 
  \item This is a turning point, as a grade A in the thermodynamics exam is generally considered proof of the outstanding intellectual abilities of the candidate.
  \item We present the results of our test and discuss the consequences of its outcome for teaching and engineering practice.
\end{itemize}


\tableofcontents  

\section{Exam on Thermodynamics (in German)}
In this section, we present the exam in thermodynamics that has been given to our students this year as well as to the model o3. The exam has been written in German.

\clearpage
\includepdf[pages=-]{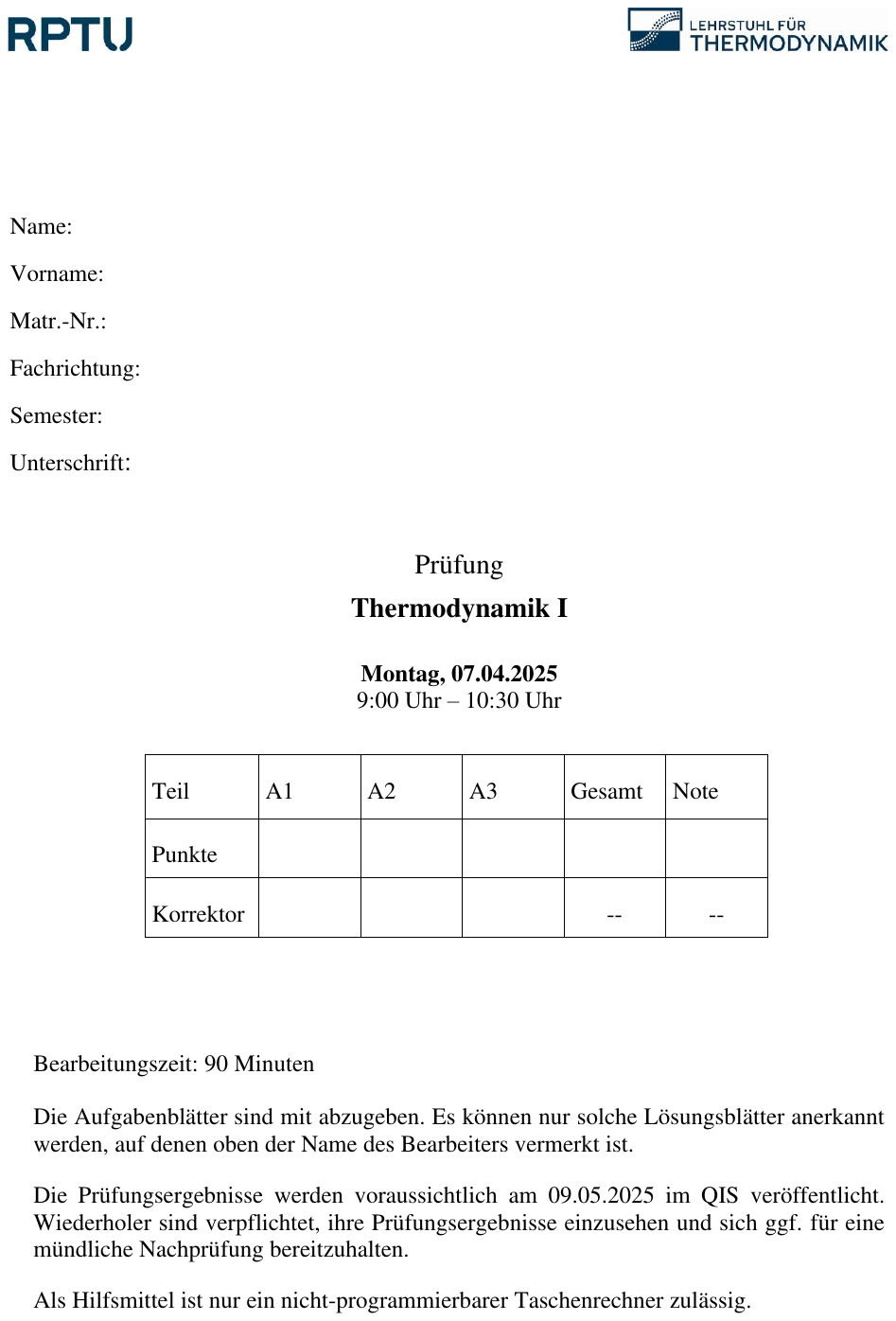}

\newpage
\section{Solution of the Exam on Thermodynamics}
In this section, we present a solution, which was used to evaluate the students as well as the o3 model on the exam in thermodynamics. 

\clearpage
\includepdf[pages=-]{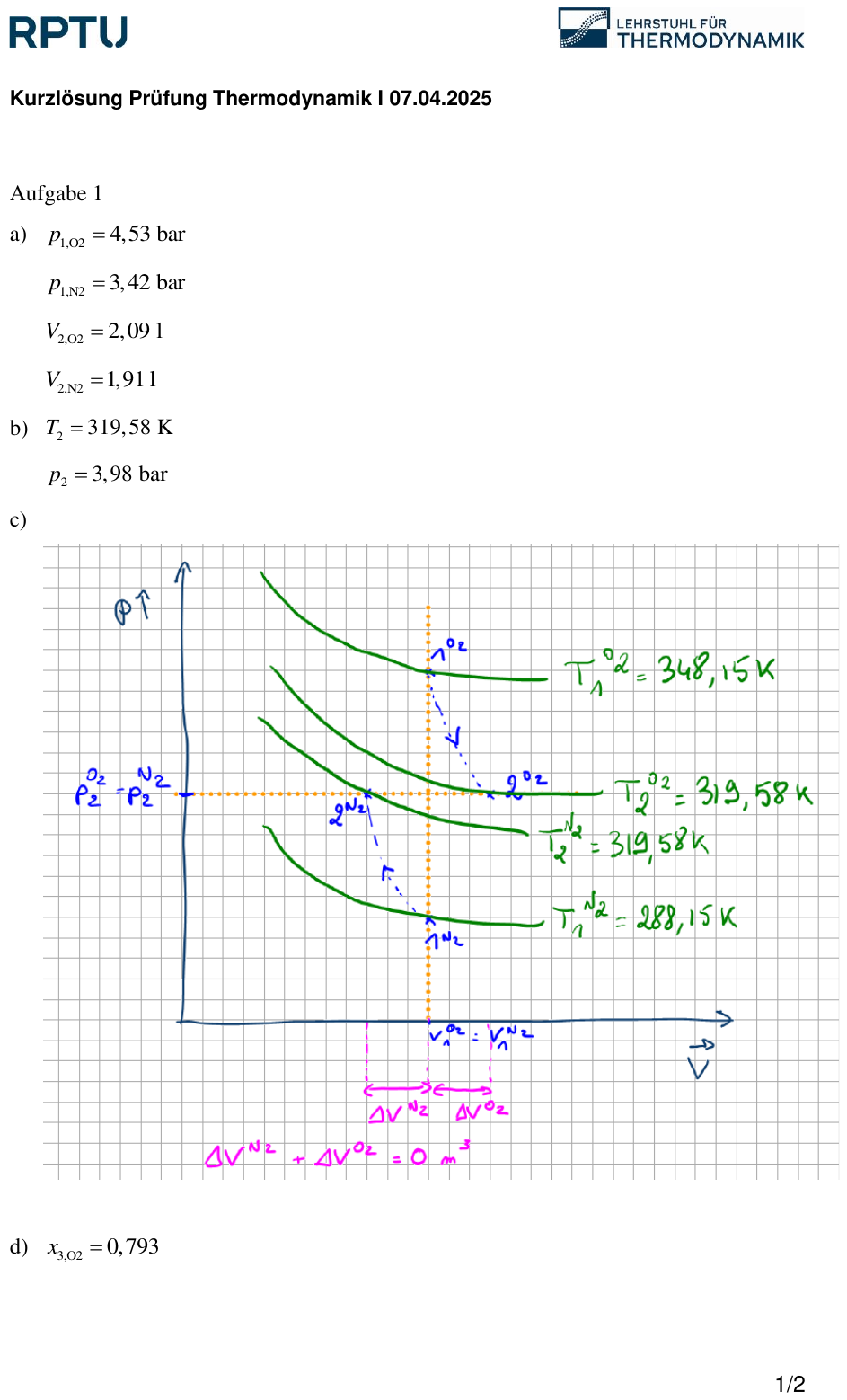}